\documentclass{article}
\usepackage{arxiv}

\usepackage[utf8]{inputenc} 
\usepackage[T1]{fontenc}    
\usepackage{hyperref}       
\usepackage{url}            
\usepackage{booktabs}       
\usepackage{amsfonts}       
\usepackage{nicefrac}       
\usepackage{microtype}      
\usepackage{lipsum}
\usepackage{amsmath,mathtools,bbm,multirow,url,hyperref,enumerate}
\usepackage{color}
\usepackage{float}
\usepackage{hyperref}
\usepackage{makecell}
\usepackage{algorithm,algorithmic}
\usepackage{multicol}
\usepackage{enumitem}

\usepackage{natbib}
\hypersetup{colorlinks,citecolor=blue}

\title{Stratified modestly-weighted log-rank tests in settings with an anticipated delayed separation of survival curves}

\author{
  Dominic Magirr \\
  Advanced Methodology and Data Science\\
  Novartis Pharma AG\\
  Basel, Switzerland\\
  \texttt{dominic.magirr@novartis.com} \\
   \And
  Jos\'e L. Jim\'enez \\
  Global Drug Development\\
  Novartis Pharma AG\\
  Basel, Switzerland\\
  \texttt{jose\_luis.jimenez@novartis.com} \\
}

\begin{document}

\maketitle

\begin{abstract}
Delayed separation of survival curves is a common occurrence in confirmatory studies in immuno-oncology. Many novel statistical methods that aim to efficiently capture potential long-term survival improvements have been proposed in recent years. However, the vast majority do not consider stratification, which is a major limitation considering that most (if not all) large confirmatory studies currently employ a stratified primary analysis. In this article, we combine recently proposed weighted log-rank tests that have been designed to work well under a delayed separation of survival curves, with stratification by a baseline variable. The aim is to increase the efficiency of the test when the stratifying variable is highly prognostic for survival. As there are many potential ways to combine the two techniques, we compare several possibilities in an extensive simulation study. We also apply the techniques retrospectively to two recent randomized clinical trials.
\end{abstract}

\section{Introduction}

The predominant method for analysing time-to-event endpoints in oncology randomized clinical trials (RCTs) is the (stratified) Cox model or log-rank test. In immuno-oncology, and in particular for trials comparing immune-checkpoint inhibitors with other forms of therapy, the Cox model has remained the default analysis choice despite a clear and consistent pattern of non-proportional hazards across studies (\cite{rahman2019deviation}). More specifically, the form of non-proportional hazards is typically a delayed separation (or perhaps even crossing) of survival curves. Although much attention has been paid to alternative forms of analysis, including weighted log-rank tests that aim to capture long-term improvements in survival, such methods are yet to be used in practice (\cite{freidlin2019methods,freidlin2020reply,huang2020estimating,uno2020log}), with few exceptions (\cite{wu2019nivolumab, kojima2020randomized}). One practical limitation is that most large phase 3 trials in oncology employ a stratified analysis in order to increase efficiency in the presence of prognostic covariates. The vast majority of recent proposals for improved analysis in the presence of non-proportional hazards ignore the issue of prognostic covariates. The goal of this paper is to establish whether or not it is possible to retain the benefits of covariate adjustment and weighted log-rank tests when they are used together in a stratified weighted log-rank test.

To motivate our investigations, we take advantage of the (de-identified) patient-level data available from the OAK (\texttt{NCT02008227}) and POPLAR (\texttt{NCT01903993}) clinical trials (see \cite{rittmeyer2017atezolizumab} and \cite{fehrenbacher2016atezolizumab}). Both (randomized) studies compare atezolizumab versus docetaxel in patients with non-small-cell lung cancer and their Kaplan-Meier curves exhibit the late separation pattern often seen with immunotherapy agents. The data from both studies is available in \cite{gandara2018blood} and, apart from the survival data, it contains data on several covariates, including the Eastern Cooperative Oncology Group (ECOG) performance status, which measures the physical capability of patients \cite{oken1982toxicity}. 

\begin{figure}[h]
\caption{Kaplan-Meier curves by strata and treatment group in the OAK trial.}
\centering
\vspace{0.25cm}
\includegraphics[scale=0.5]{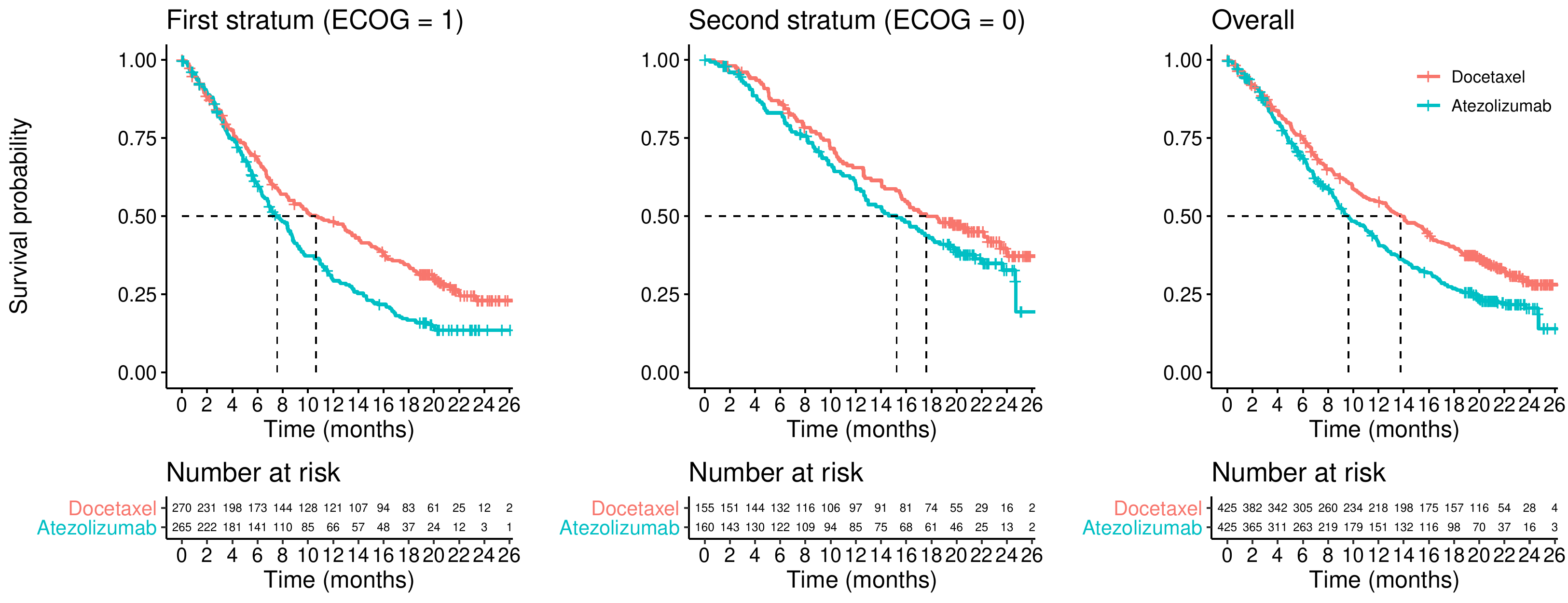}
\label{oak_plots}
\end{figure}

\begin{figure}[h]
\caption{Kaplan-Meier curves by strata and treatment group in the POPLAR trial.}
\centering
\vspace{0.25cm}
\includegraphics[scale=0.5]{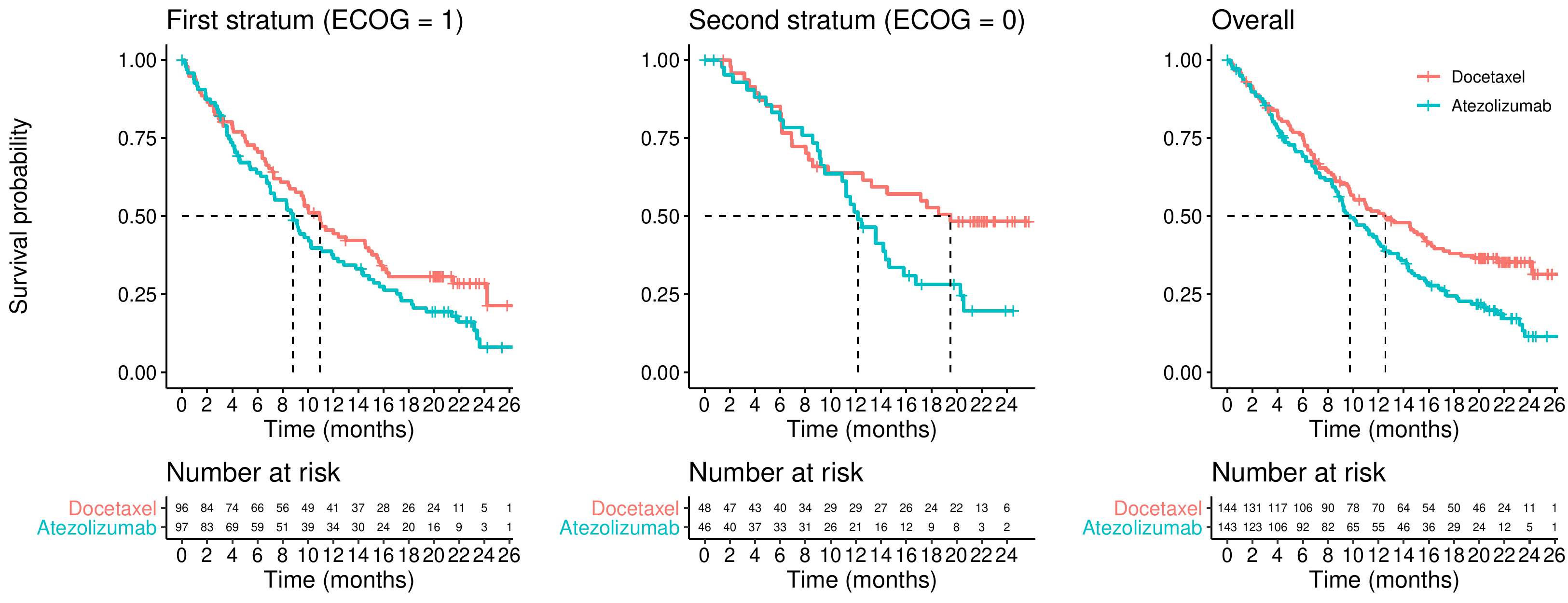}
\label{poplar_plots}
\end{figure}

Figures \ref{oak_plots} and \ref{poplar_plots} show the overall survival for the two treatment groups in the OAK and POPLAR trials, respectively, stratified by ECOG status. One can see that there is a delay of several months before the separation of the survival curves. This feature is not unique to OAK and POPLAR as it has been observed in many trials involving immune-checkpoint inhibitors (see e.g., \cite{owen2021delayed}), and arguably, therefore, is predictable at the design stage. A second feature of these figures is that patients with baseline ECOG=0 have, on average, longer survival times than patients with ECOG=1 at baseline. 

These two features (i.e., a prognostic covariate and a delayed separation of survival curves) motivate the rest of this paper. In isolation, both features have been studied extensively in the literature. Regarding the importance of adjusting for prognostic covariates, we refer the reader to \cite{hauck1998should,mehrotra2012efficient,xu2019hazard}. Regarding the use of weighted log-rank tests to increase power in situations with non-proportional hazards, we refer the reader to \cite{fleming2011counting, jimenez2019properties, magirr2019modestly,lin2020alternative,jimenez2020quantifying}. In this paper, we wish to determine a good analysis strategy when both of these two features can be anticipated at the design stage. To that end, we shall investigate multiple combinations of stratified and weighted log-rank tests.

The article is structured as follows. In section \ref{sc_stratified_logrank}, we introduce the basic theory in order to build both the stratified log-rank and stratified weighted log-rank tests. In section \ref{sc_simulation_study}, we present a simulation study, where we thoroughly explore the influence of a prognostic covariate on the power, varying both its prognostic strength and degree of treatment effect modification. In Section \ref{sc_case_studies}, we describe the OAK and POPLAR data in more detail and  apply a series of stratified and/or weighted log-rank tests for illustration. We conclude with a discussion and recommendations in section \ref{sc_discussion}.

\section{Stratified log-rank test}
\label{sc_stratified_logrank}

Our general strategy for constructing stratified weighted log-rank tests is to mimic the structure of the stratified log-rank test. We shall therefore review this test for the specific case of two strata. Extensions to more strata are conceptually straightforward. Throughout this manuscript we are assuming a very limited number of strata, perhaps between two and six, say, with no issues regarding sparse data. This is likely reasonable for confirmatory oncology studies. 

The null hypothesis that is generally considered when performing a stratified log-rank test is:
\begin{equation}
\label{eq_null_hypothesis}
    H_{0,S}: S_{E,i}(t) = S_{C,i}(t)\text{ for all }t>0\text{ and for all strata }i
\end{equation}
where $S_{E,i}$ and $S_{C,i}$ denote the survival distributions on the experimental and control arms, respectively, in the $i$th stratum $(i = 1,2)$. Though less conventional, one could also consider a one-sided null hypothesis:
\begin{equation}
    \tilde{H}_{0,S}: S_{E,i}(t) \leq S_{C,i}(t)\text{ for all }t>0\text{ and for all strata }i.
\end{equation}
Suppose in each stratum ($i=1,2$) we have ordered distinct event times $t_j$ ($j = 1,\ldots, d_i$). The test statistic is derived by constructing $d_1 + d_2$ 2x2 tables such as Table \ref{two_by_two} depicting the data from stratum $i$ at time $t_j$. Conditional on the margins of each 2x2 table, and assuming identical survival curves on the two treatment arms within strata, the observed number of events on the experimental treatment at event time $t_j$ in stratum $i$, denoted by $O_{i,1,j}$, follows a hypergeometric distribution, where the expected number of events is $E_{i,1,j} = O_{i,j}\times n_{i,1,j} / n_{i,j}$, and the variance of $O_{i,1,j}$ is
\begin{equation*}
 V_{i,1,j} = \frac{n_{i,0,j}n_{i,1,j}O_{i,j}(n_{i,j} - O_{i,j})}{n_{i,j}^2(n_{i,j} - 1)}   
\end{equation*}

\begin{table}[ht]
\caption{A 2x2 table describing the situation at event time $t_j$ in stratum $i$.}
\centering
\begin{tabular}{c|cc|c}
  & Event = Y & Event = N &   \\ 
  \hline
Trt = 1 & $O_{i,1,j}$ & $n_{i,1,j} - O_{i,1,j}$ & $n_{i,1,j}$ \\ 
  Trt = 0 & $O_{i,0,j}$ & $n_{i,0,j} - O_{i,0,j}$ & $n_{i,0,j}$ \\ 
   \hline
  & $O_{i,j}$ & $n_{i,j} - O_{i,j}$ & $n_{i,j}$ \\ 
  \end{tabular}
  \label{two_by_two}
\end{table}

It can be shown that, asymptotically,

\begin{equation}
\tilde{U} = U_1 + U_2 \sim N(0, V_1 + V_2)
\end{equation}

where $U_i = \sum_{j = 1}^{d_i} \left( O_{i,1,j} - E_{i,1,j}\right)$ and $V_i = \sum_{j = 1}^{d_i}V_{i,1,j}$. We shall let $\tilde{Z}:= \tilde{U}/\sqrt{V_1+V_2}$ denote the standardized Z-statistic for stratified log-rank test.

\subsection{Stratified weighted log-rank test}

If we only had one stratum, then we could construct a weighted log-rank test statistic as
\begin{equation}
    U_i^W:= \sum_{j = 1}^{d_i} w_{i,j}  \left( O_{i,1,j} - E_{i,1,j}\right) \sim N(0, V_i^W)
\end{equation}
where $w_{i,j}$ are a choice of weights, and $V_i^W = \sum_{j = 1}^{d_i}w_{i,j}^2V_{i,1,j}$.

When we have two strata, we need to combine $U_1^W$ and $U_2^W$ to create an overall test statistic. There are many ways one could do this. We shall investigate three options:

\begin{enumerate}[label=\roman*)]
\item Mimic the stratified log-rank test directly by taking the sum of the score statistics,

$$\tilde{U}^{W,u}:= U_1^W + U_2^W \sim N(0, V_1^W + V_2^W),$$

which we shall standardize to $\tilde{Z}^{W,u}$.

\item Express the stratified log-rank test as a linear combination of standardized Z-statistics, and then replace the Z-statistics with the standardized weighted log-rank statistics,

\begin{equation}
   \tilde{U}^{W,z}:= \sqrt{V_1}\left(\frac{U_1^W}{\sqrt{V_1^W}}\right) + \sqrt{V_2}\left(\frac{U_2^W}{\sqrt{V_2^W}}\right) \sim N(0, V_1 + V_2) 
\end{equation}which we shall standardize to $\tilde{Z}^{W,z}$.

\item Combine the log-hazard ratios according to the sample size in each stratum ($n_i$) as proposed by \cite{mehrotra2012efficient}. We modify their approach slightly by replacing stratum specific maximum likelihood estimates with stratum specific Peto estimates (\cite{berry1991comparison}) for the log-hazard ratio. The test statistic can be written as
\begin{equation}
  \tilde{\theta}^{n} \propto n_1\left(\frac{U_1}{{V_1}}\right) + n_2\left(\frac{U_2}{V_2}\right) \sim N\left(0, \frac{n_1^2}{V_1} + \frac{n_2^2}{V_2} \right).  
\end{equation}

We therefore propose to replace $U_i/V_i$ with $U^W_i/V^W_i$, giving
\begin{equation}
\tilde{\theta}^{W,n} \propto n_1\left(\frac{U^W_1}{{V^W_1}}\right) + n_2\left(\frac{U^W_2}{V^W_2}\right) \sim N\left(0, \frac{n_1^2}{V^W_1} + \frac{n_2^2}{V^W_2} \right),
\end{equation}
which we shall standardize to $\tilde{Z}^{W,n}$.

\end{enumerate}

The first method, $\tilde{Z}^{W,u}$, might appear the most obvious way to combine strata based on the structure of the stratified log-rank test. However, there is heuristic justification for $\tilde{Z}^{W,z}$, in the sense that $V_1$ and $V_2$ are approximately proportional to the number of events in each stratum. Intuitively, the number of events might be a reasonable measure of how much information each stratum is contributing, even under non-proportional hazards. Also, Mehrotra et al. have reported good performance of $\tilde{\theta}^{n}$, which may translate to $\tilde{Z}^{W,n}$.

In this article we focus on the modestly-weighted log-rank test proposed by \cite{magirr2019modestly}, although other weights could be easily used if deemed more appropriate by the clinical team. The weights of the modestly-weighted log-rank test that we use in this article are calculated as $w_j = 1 / \max{\left\lbrace \hat{S}(t_j-), \hat{S}(t^*)\right\rbrace}$, where $\hat{S}$ denotes the Kaplan-Meier estimate from the pooled sample. The modestly-weighted test can be thought of as similar to an average landmark analysis from time $t^*$ to the end of follow up \cite{magirr2021non,magirr2021design}. An important point is that if we anticipate a delay of, for example, 6 months, this does not  necessarily mean that we should choose $t^*=6$. A somewhat later $t^*$ (e.g., $t^* = 12$) will tend to have higher power since it gives chance for the curves to separate. On the other hand, if there is some uncertainty regarding the length of the delay, then choosing a value of $t^*$ closer to zero protects the power in case the proportional hazards assumption holds. It is important to mention that $t^*=0$ is the same as the standard log-rank test and that, for values of $t^*$ close to $0$, there will be little difference between these two tests. For further discussion of the choice of $t^*$ and the choice of weights more generally, we refer the reader to the papers \cite{magirr2019modestly,magirr2021design}. Also bear in mind that, while there are differences in terms of performance between the modestly-weighted log-rank test and the use of other weighted tests (e.g., the Fleming and Harrington class of weights (\cite{fleming2011counting})), it is not in the scope of this article to make such a comparison. We refer the reader to previous publications making these comparisons (\cite{magirr2019modestly,magirr2021non}), as well as others discussing the properties of alternative weighting schemes (\cite{fleming2011counting,yang2010improved,gares2014comparison,karrison2016versatile,roychoudhury2021robust,jimenez2019properties}).  

\section{Simulation study}
\label{sc_simulation_study}

We propose to evaluate the following test statistics:

\begin{itemize}
\item $Z$: unstratified log-rank test
\item $Z^W$: unstratified weighted log-rank test 
\item $\tilde{Z}$: stratified log-rank test
\item $\tilde{Z^n}$: stratified log-rank test (Mehrotra et al.)
\item $\tilde{Z}^{W,u}$: stratified weighted log-rank test (U-statistic scale)
\item $\tilde{Z}^{W,z}$: stratified weighted log-rank test (Z-statistic scale)
\item $\tilde{Z}^{W,n}$: stratified weighted log-rank test (sample size scale)
\end{itemize}

For the modestly-weighted log-rank test we shall use $t^* = 12$ months. As discussed above, this may be reasonable when survival curves are anticipated to separate at around $t = 6$. We choose not to test other values of $t^*$ in this simulation study since its purpose is not to assess the sensitivity of the modestly-weighted log-rank test to the choice of $t^*$, which has already been shown to be robust (see \cite{magirr2019modestly,magirr2021non,ghosh2021robust}).

It should be emphasized that the unstratified tests are usually testing the null hypothesis $H_0: S_{E}(t) = S_{C}(t)\text{ for all }t>0$ in the full trial population. However, because $H_{0,S}$ is contained in $H_0$, one could use $Z$ or $Z^W$ to test $H_{0,S}$. The reverse is not true, and it is not generally valid to test $H_0$ using any of the stratified test statistics.

The range of scenarios that we shall consider have been designed with the following features in mind which are all predictable from theory:

\begin{itemize}
\item $\alpha$ control for $H_{0,S}$ for all tests.
\item $\alpha$ control for $H_{0}$ using $U$ and $U^W$, but not necessarily if one were to (incorrectly) use a stratified test statistic.
\item Superior power of stratified tests compared to unstratified tests when stratifying on a strong prognostic covariate.
\item Superior power of weighted tests over unweighted tests under non-proportional hazards (delayed separation) scenarios, and vice-versa under proportional hazards.
\end{itemize}

Beyond highlighting such features, the second aim of the simulation study is to highlight the trade-offs involved between the various stratified-weighted tests described above.  A third goal is to explore the relative importance of stratifying versus weighting. In this respect we urge caution, however, since it is only possible to explore a relatively small number of situations, and conclusions should not necessarily be extrapolated beyond the scenarios considered here. 

\subsection{Basic trial design}

We consider a basic trial design with total duration of 24 months, including a recruitment period of 9 months. We assume that 344 patients are enrolled at a uniform rate. The randomization ratio is 1:1. The censoring distribution is driven purely by the recruitment times and the end of the study. We do not assume any other censoring. The sample size was arrived at by considering base case survival distributions in the full population under experimental treatment and control (\cite{schoenfeld1983sample}), namely an exponential distribution with median 8 months and 12 months, respectively. With a one-sided $\alpha$ of 2.5\% and power 90\% this would require a total of $4 \times \left ( \frac{\Phi^{-1}(0.9) + \Phi^{-1}(0.975)}{-\log(8/12)} \right )^2 \approx 256$ events. Based on the recruitment assumptions and desired total trial duration, this would require 344 patients, 172 in each arm. Note that no stratifying variables are considered in this sample size calculation, as is typically the case. Also, we have used an exponential assumption only to get the sample size in the right ballpark. In practice, under the assumption of delayed separation, a simulation study to empirically estimate the power under a realistic (based on the prior knowledge) time-to-event distribution (e.g., piecewise exponential) may be necessary, especially if the scenario is complex. Then, if the power is too low, one could simply tweak either the sample size or the follow-up period as needed.

\subsection{Simulation scenarios}
\label{sc_simulation_scenarios}

We shall assume that there is a single binary baseline covariate of interest, so that there are only two strata, each with a prevalence of 50\%. A realistic example would be an ECOG performance status 0 or 1. Individuals with ECOG = 0 are fully active whereas individuals with ECOG = 1 have some sort of physical limitation. We shall consider three scenarios for the survival distributions in the two strata on the control arm:

\begin{itemize}
  \item Non-prognostic: The survival distribution in each stratum is the same as the base case distribution (i.e., an exponential distribution with a median of 8 months).
  
  \item Moderate prognostic: The survival distributions in the first (ECOG = 1) and second (ECOG = 0) strata are exponential with medians of 6 months and 10 months, respectively. In other words, the stratum ECOG = 1 has slightly lower survival than the stratum ECOG = 0.
  
  \item Strong prognostic: The survival distributions in the first (ECOG = 1) and second (ECOG = 0) strata are exponential with medians of 3 months and 15 months, respectively. In other words, the stratum ECOG = 1 has a much lower survival than the stratum ECOG = 0.
\end{itemize}

\begin{figure}[H]
\caption{Scenarios under proportional hazards.}
\centering
\vspace{0.25cm}
\includegraphics[scale=0.7]{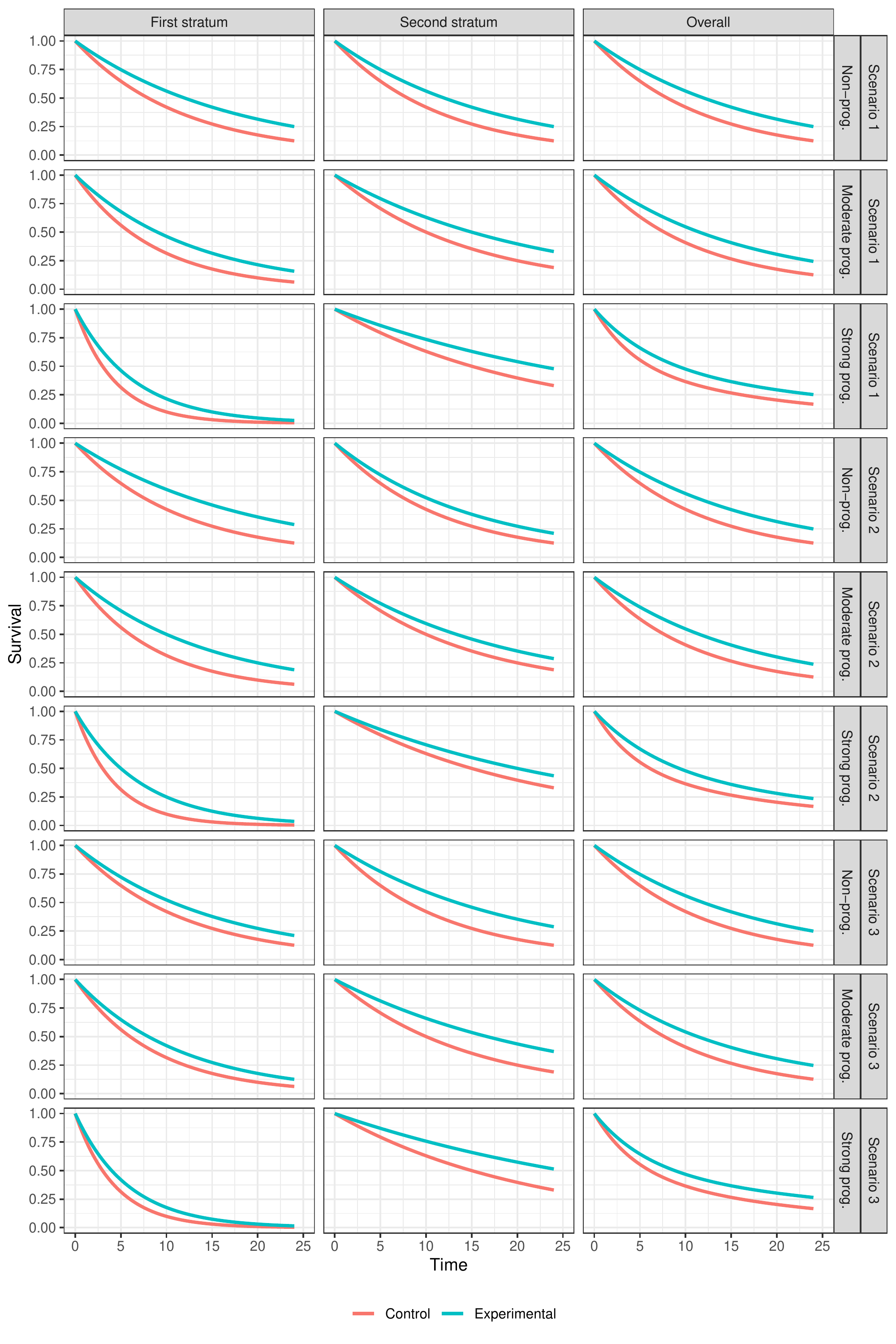}
\label{plot_ph}
\end{figure}

\begin{figure}[H]
\caption{Scenarios under null effect.}
\centering
\vspace{0.25cm}
\includegraphics[scale=0.7]{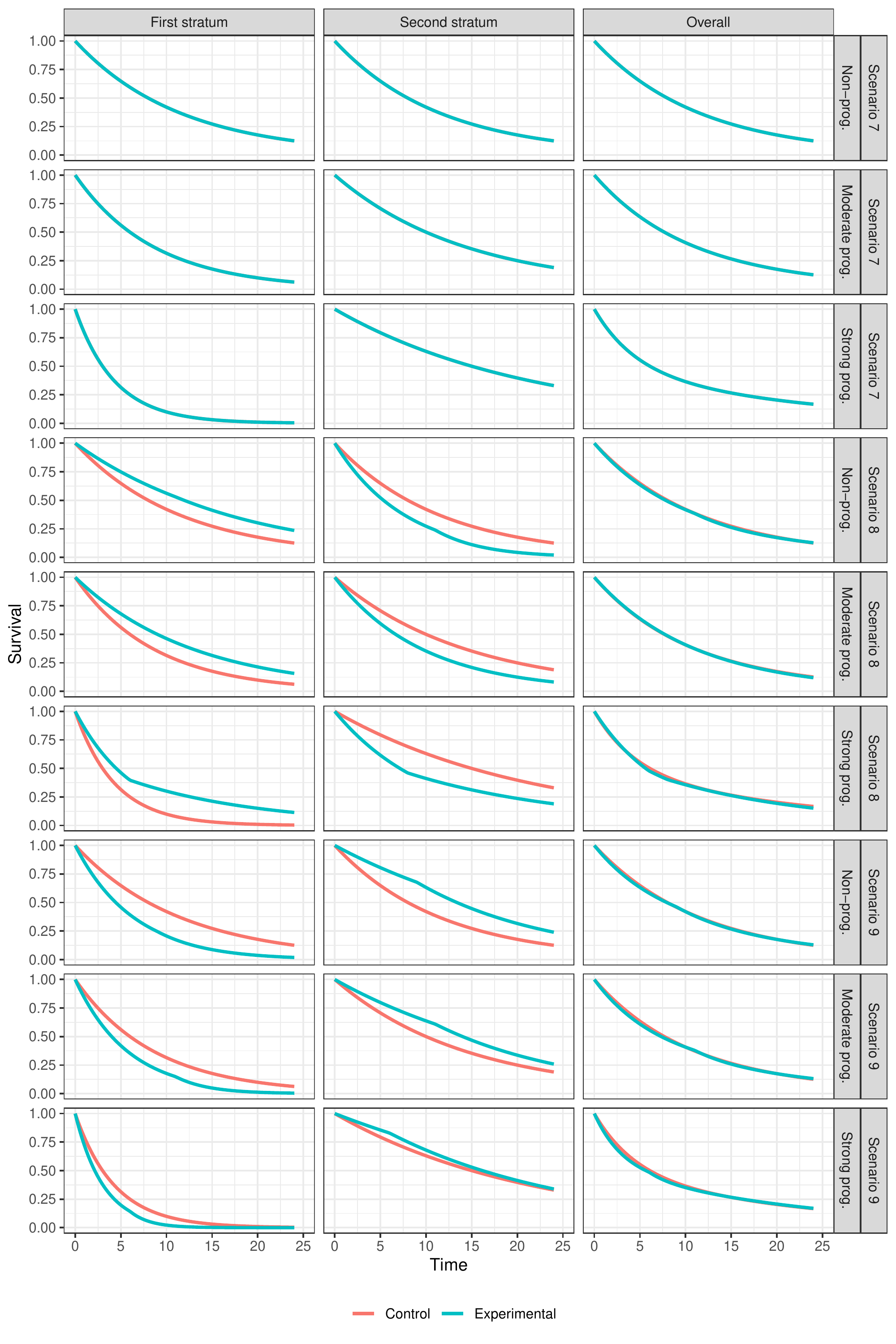}
\label{plot_null_effect}
\end{figure}

Apart from prognostic strength, we shall also consider the following (nine) scenarios for treatment effect modification:

\begin{itemize}

\item Scenarios under proportional hazards (presented in Figure \ref{plot_ph}):

\begin{itemize}
    \item Scenario 1: Scenario with homogeneous survival distributions across strata (i.e., we have the same hazard ratio in each stratum).
    \item Scenario 2: Scenario in which the stratum with poor prognosis has a better (lower) hazard ratio than the stratum with good prognosis.
    \item Scenario 3: Scenario in which the stratum with poor prognosis has a worse (larger) hazard ratio than the stratum with good prognosis.
\end{itemize}

\item Scenarios under delayed survival curve separation (presented in Figure \ref{plot_nph}):

\begin{itemize}
\item Scenario 4: Scenario with homogeneous delayed separation of survival curves, where the delay is of equal length in each stratum. The difference in survival probability (i.e., experimental - control) at late follow-up times is also similar in the two strata.
\item Scenario 5: Scenario in which the stratum with poor prognosis has a better second-period treatment effect than the stratum with good prognosis. There is also an equal delay in each stratum.
\item Scenario 6: Scenario in which the stratum with poor prognosis has a worse second-period treatment effect than the stratum with good prognosis. There is also an equal delay in each stratum.
\end{itemize}

\item Scenarios with null overall effect (presented in Figure \ref{plot_null_effect}):

\begin{itemize}
    \item Scenario 7: Scenario with hazard ratio of 1 in each stratum as well as overall.
    \item Scenario 8: Scenario in which there is positive difference in survival probability (i.e., experimental better than control) in the poor prognosis stratum and a negative  difference in survival probability (i.e., experimental worse than control) in the good prognosis stratum. The overall hazard ratio is equal to 1, approximately.
    \item Scenario 9: Scenario in which there is negative difference in survival probability (i.e., experimental worse than control) in the poor prognosis stratum and a positive difference in survival probability (i.e., experimental better than control) in the good prognosis stratum. The overall hazard ratio is equal to 1, approximately.
\end{itemize}

\end{itemize}

In total, combining the three options for the prognostic effect with the nine options for the treatment effects, we have $3 \times 9 = 27$ different scenarios to evaluate that we display graphically in the supplementary appendix. Scenarios 2 and 3, and scenarios 5 and 6 are essentially the same (symmetric) for the non-prognostic covariate case, so strictly speaking there are only 25 unique scenarios, but we ignore this for ease of presentation.

\subsection{Results}
In Figure \ref{plot_tests}, we present the estimated power in each of the 27 scenarios described in section \ref{sc_simulation_scenarios}.

\subsubsection{Null scenarios}

\begin{figure}[t]
\caption{Power and type-I error using different stratified and unstratified log-rank and weighted log-rank tests. The upper and lower dashed represent the value 0.9 and 0.025, respectively. The heading "poor prog. '>' good prog." indicates that the stratum with the poor prognosis has a better treatment effect than the stratum with good prognosis. }
\centering
\vspace{0.25cm}
\includegraphics[scale=0.65]{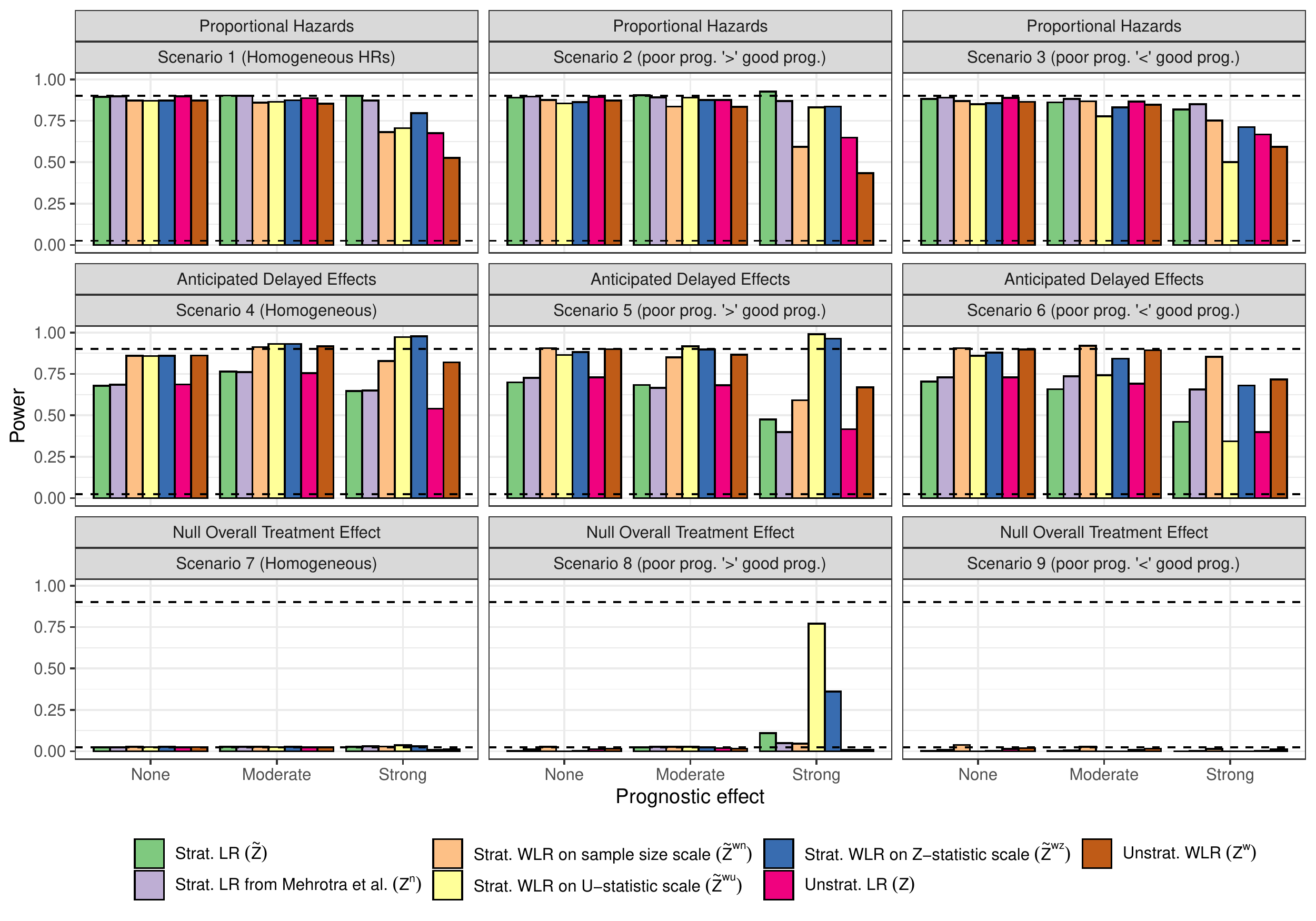}
\label{plot_tests}
\end{figure}

The most striking result in Figure \ref{plot_tests} is the "power" of the stratified tests when the marginal treatment effect is zero but there is a strong prognostic effect, with the poor prognosis stratum having treatment benefit, and the better prognosis stratum receiving a harmful treatment effect. In this scenario, for all the stratified tests, the power is above 2.5\%, sometimes massively so. 

Two observations help to describe this phenomenon. Firstly, we can think about the stratified tests as a weighted combination of two standardized z-statistics, one from each stratum. Even when this weighting is equal (as is the case here for $\tilde{Z}^{W,n}$ and $\tilde{Z}^{n}$), the z-statistic coming from the stratum with poor prognosis will be based on many more events than the z-statistic from the better prognosis stratum. Even if the magnitude of treatment effect is similar (but of opposite direction) in the two strata, the poor prognosis stratum will tend to have a standardized z-statistic with a non-centrality parameter that is larger in magnitude than that of the better prognosis stratum. This explains why the power for $\tilde{Z}^{n}$ and $\tilde{Z}^{W,n}$ is above 2.5\%. It is a consequence of censoring: the better prognosis stratum is more heavily censored (less mature) than the poor prognosis stratum. For other types of non-censored data analysed via generalized linear models with canonical link functions, this phenomenon does not occur, and stratified test statistics would control alpha also for the marginal null hypothesis, at least asymptotically (see \cite{rosenblum2016matching}).

The second observation is that as we move from $\tilde{Z}^{n}$ to $\tilde{Z}$, and from $\tilde{Z}^{W,n}$ to $\tilde{Z}^{W,z}$ to $\tilde{Z}^{W,u}$, the relative weighting of the within-strata z-statistics gets more and more extreme in favour of the stratum with more events. This explains the difference in power between the various stratified weighted tests. 

When it is the better prognosis stratum that has the treatment benefit, the reverse happens, and the power of the stratified tests is well below 2.5\% (bottom right panel of Figure \ref{plot_tests}).

\subsubsection{Proportional hazards scenarios}

As one would expect, the stratified log-rank test ($\tilde{Z}$) and Mehrotra et al's stratified log-rank test ($\tilde{Z}^{n}$) are the most powerful methods, with the difference compared to unstratified tests most apparent when there is a strong prognostic effect. 

The stratified weighted log-rank tests remain competitive with the stratified log-rank tests, even under proportional hazards, as long as the prognostic effect is moderate. However, for large prognostic effects, some substantial differences emerge. 

Among the subset of stratified weighted log-rank tests, there is no uniformly most (or least) powerful method. It depends on whether the poor prognosis stratum has a larger treatment effect than the better prognosis stratum, or vice versa. The test statistic that gives the most extreme  weight to the stratum with more events "wins" in one scenario and "loses" in the other. This is entirely analogous to the discussion above for the null scenarios. Something similar happens if we compare $\tilde{Z}$ with $\tilde{Z}^{n}$: the test that gives a more extreme weight in favour of the stratum with more events ($\tilde{Z}$) performs better when that stratum also has the larger treatment effect, and vice versa when the stratum with more events has a smaller treatment effect. When the hazard ratio is equal in both strata then the standard stratified log-rank test is optimal, of course.

\subsubsection{Non-proportional hazards scenarios}

When comparing power under scenarios with a delayed separation of survival curves, much of the above discussion about the relative merits of the test statistics still applies. In this case, however, the weighted test statistics perform much better than the unweighted versions, as one would expect.

Based on this particular simulation study, it appears that using a weighted log-rank test in anticipation of a delayed separation has a greater impact on power than using a stratified test in anticipation of a prognostic effect. However, this is only a small selection of scenarios, and something that should be judged on a case-by-case basis.

Of the tests that combine weighting and stratification, perhaps the combined test on the Z-scale ($\tilde{Z}^{W,z}$) is most attractive. Since it weights the strata in a manner that is intermediate out of $\tilde{Z}^{W,n}$, $\tilde{Z}^{W,z}$ and $\tilde{Z}^{W,u}$, it appears to be most robust to the strength and direction of stratum-specific treatment effects. It performs consistently well across all the chosen scenarios.

\section{Cases studies: The OAK and POPLAR trials}
\label{sc_case_studies}

As previously mentioned, the OAK and POPLAR trials are two (randomized) studies that compare atezolizumab versus docetaxel in patients with non-small-cell lung cancer. Both trials exhibit a late separation pattern of their survival curves. In Figures \ref{oak_plots} and \ref{poplar_plots} we present the first stratum, second stratum and overall Kaplan-Meier curves from the OAK and POPLAR trials, respectively.

\begin{figure}[H]
\caption{Scenarios under non-proportional hazards.}
\centering
\vspace{0.25cm}
\includegraphics[scale=0.7]{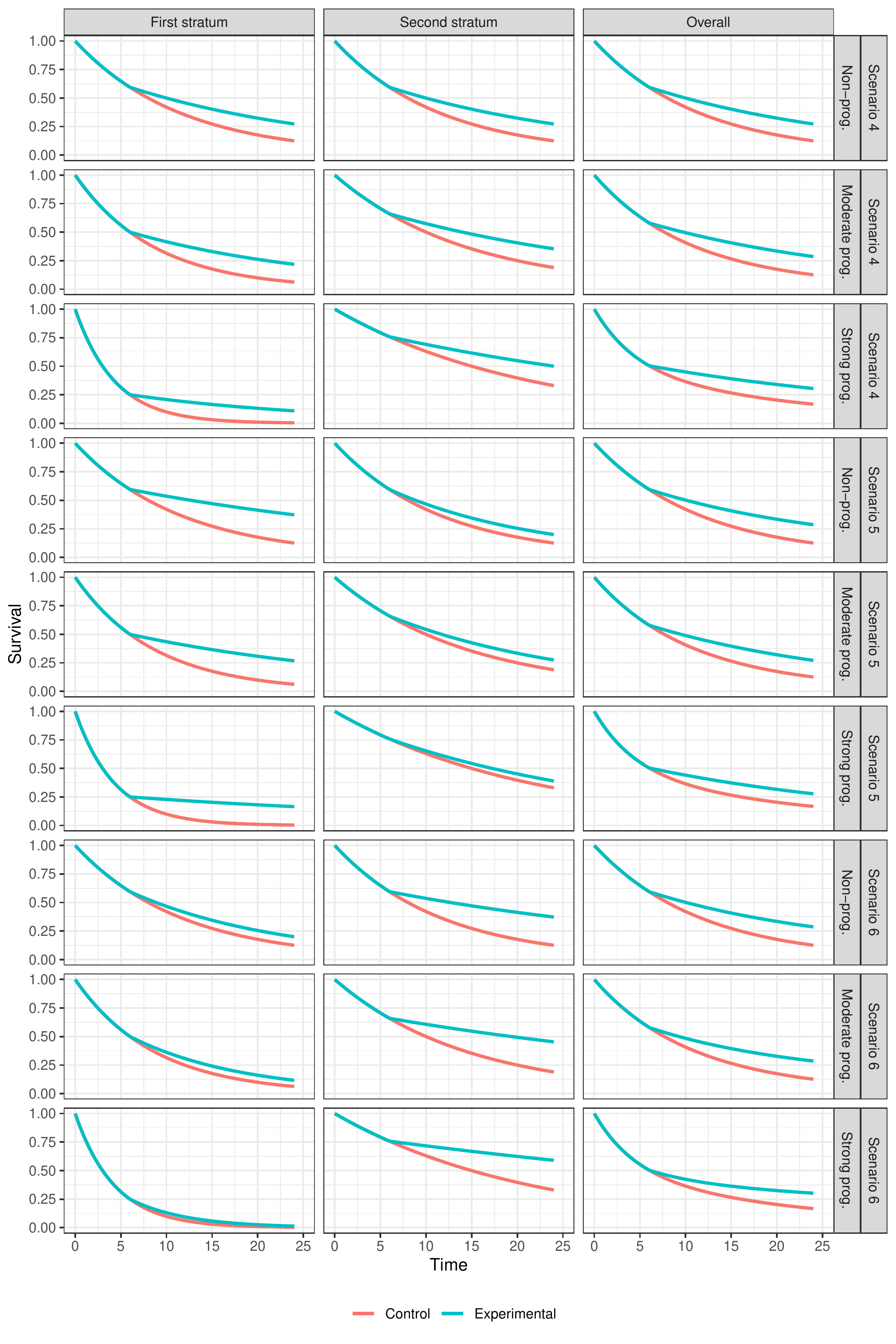}
\label{plot_nph}
\end{figure}

In the OAK trial the survival curves in the first stratum are moderately lower than those from the second stratum (i.e., there is a moderate prognostic effect). We also see that the treatment effect in the stratum with worse prognosis appears larger than the one from the stratum with better prognosis. These characteristics are similar to those evaluated in scenario 5 (see Figure \ref{plot_nph}, scenario 5, moderate prognosis).

In the POPLAR trial the survival curves in the first stratum are also somewhat lower than those from the second stratum, but the prognostic effect is less pronounced than for the OAK trial. In this case, we see that the treatment effect in the stratum with worse prognosis appears smaller than the one from the stratum with better prognosis. These characteristics are similar to those evaluated in scenario 6 (see Figure \ref{plot_nph}, scenario 6, for moderate (or no) prognostic effect).

For illustration purposes, we implement the stratified and unstratified tests described in section \ref{sc_simulation_study} using the OAK and POPLAR trials' data. The test statistics are presented in Table \ref{table_tests_oak_poplar}. Note that with the OAK and POPLAR datasets, we use $t^* = 6$ and $t^* = 12$ in the modestly-weighted log-rank test. The reason is that Magirr and Jim\'enez (2021) already used $t^* = 6$ in the POPLAR trial and so, for consistency with both the existing literature and the simulation study we present in this article, we show the values of the tests statistics with both $t^* = 6$ and $t^* = 12$.


With the OAK trial data, we observe that the stratified weighted log-rank test (combined on the U-statistic scale) has the best z-statistic (smallest p-value). This is consistent with what we would expect based on the simulation study. 

With the POPLAR trial data, we observe that the unstratified weighted log-rank test has the best Z statistic (smallest p-value). This may be partly explained by the smaller observed prognostic effect of ECOG status in the POPLAR trial as compared to OAK.

Overall, we observe that the weighted test statistics are lower with $t^* = 12$ than with $t^* = 6$. This is sensible and due to the fact that survival curves start to split at approximately $t = 4$ and thus $t^* = 6$ is perhaps too close to that moment where the full treatment effect is not yet fully observable.

\begin{table}[]
\caption{Tests statistics obtained from the OAK and POPLAR datasets using $t^* = 6$ and $t^* = 12$.}
\centering
\vspace{0.25cm}
\begin{tabular}{c|cc|cc}
\multirow{2}{*}{Test Statistics} & \multicolumn{2}{c|}{OAK Trial} & \multicolumn{2}{c}{POPLAR Trial} \\ 
 & \multicolumn{1}{c}{$t^* = 6$} & $t^* = 12$ & \multicolumn{1}{c}{$t^* = 6$} & $t^* = 12$ \\ \hline
$Z$ (unstratified log-rank test) & \multicolumn{1}{c}{-3.78} & -3.78 & \multicolumn{1}{c}{-2.77} & -2.77 \\ 
$Z^W$ (unstratified weighted log-rank test) & \multicolumn{1}{c}{-3.87} & -3.89 & \multicolumn{1}{c}{\textbf{-2.86}} & \textbf{-3.17} \\ 
$\tilde{Z}$ (stratified log-rank test) & \multicolumn{1}{c}{-4.02} & -4.02 & \multicolumn{1}{c}{-2.64} & -2.64 \\ 
$\tilde{Z^n}$ (stratified log-rank test (Mehrotra et al.)) & \multicolumn{1}{c}{-3.93} & -3.93 & \multicolumn{1}{c}{-2.73} & -2.73 \\ 
$\tilde{Z}^{W,u}$ (stratified weighted log-rank test (U-statistic scale)) & \multicolumn{1}{c}{\textbf{-4.27}} & \textbf{-4.35} & \multicolumn{1}{c}{-2.65} & -2.83 \\
$\tilde{Z}^{W,z}$ (stratified weighted log-rank test (Z-statistic scale)) & \multicolumn{1}{c}{-4.20} & -4.27 & \multicolumn{1}{c}{-2.71} & -2.94 \\ 
$\tilde{Z}^{W,n}$ (stratified weighted log-rank test (sample size scale)) & \multicolumn{1}{c}{-3.94} & -3.95 & \multicolumn{1}{c}{-2.83} & -3.13 \\ 
\end{tabular}
\label{table_tests_oak_poplar}
\end{table}

\section{Discussion}
\label{sc_discussion}

Over recent years, with the development of immunotherapies, many novel statistical methods have been proposed that aim to robustly capture potential long-term survival improvement following a delayed separation of survival curves. These methods focus, mostly, on differences in marginal survival probabilities in the full population. Stratified testing has received very little attention despite the fact that it is present in practically every large confirmatory trial.

Motivated by the OAK and POPLAR trials, two randomized studies with a delayed survival curve separation that compare atezolizumab versus docetaxel in patients with non-small-cell lung cancer, we have proposed stratified versions of weighted log-rank tests, and evaluated their properties in an extensive simulation study, taking into consideration different prognostic levels as well as the possibility of different treatment effects across strata. 

The results of the research are largely consistent with our expectations given that, when studied separately, we know that stratification leads to greater efficiency under strong prognostic effects, and (carefully chosen) weighted log-rank tests lead to greater efficiency under delayed separations of survival curves. Unsurprisingly, combining the two methods leads to the greatest efficiency when strong prognostic effects and a delay in the separation of survival curves are both present. However, we have also shown that the precise manner in which the two techniques are combined can have a large impact on power. Based on our simulation study, together with theoretical understanding, our recommendation is to use a stratified weighted log rank test where the strata are combined on the z-statistic scale. This method has robust power performance across all the scenarios we considered. 

It should be borne in mind that stratified tests correspond to the stratified null hypothesis $H_{0,S}$, and not the marginal null hypothesis $H_0$. We have shown that it is possible to construct scenarios where there is no treatment effect marginally in the full population (but there is a heterogeneous treatment effect across strata) and the power for testing $H_{0,S}$ according to an $\alpha$ level test is very much higher than $\alpha$. To put this in context, however, the situations where this happens are extreme. If considered at all plausible a-priori, one would not embark on such a trial in the full population. In addition, it is standard practice to report results for the full population (in the form of a Kaplan-Meier plot, for example), as well as separately by subgroup (for example using a forest plot). The chance of being grossly misled by a stratified test appears small. Related to this discussion is the issue of treatment effect estimation. Already, when considering non-proportional hazard situations for a homogeneous population, it is considered impossible to fully capture the treatment effect using a single number (\cite{zhao2016restricted}), and instead it is recommended to report a spectrum of treatment effects (\cite{roychoudhury2021robust}), such as differences in quantiles, milestone survival probabilities, and restricted mean survival times (\cite{uno2014moving}). When adding a heterogeneous population into the mix, the task of capturing the treatment effect adequately in a single number becomes doubly impossible. The only pragmatic way forward is to present a range of Kaplan-Meier type plots and summary measures, for a range of relevant (sub)populations. In this context, the value of the stratified weighted log-rank test statistic is simply to allow a pre-specified null hypothesis test that can produce a valid p-value, allowing a first line of defence against being misled by randomness. This should be considered one small, but important, part of the overall design and analysis of the experiment. A range of techniques are required to achieve a reasonable inference.

\bibliographystyle{plainnat}
\bibliography{references.bib}

\section*{Appendix}

\subsubsection*{Software}

The repository \texttt{\href{https://github.com/dominicmagirr/stratified_weighted_log_rank_test}{github.com/dominicmagirr/stratified\_weighted\_log\_rank\_test}} contains all R code to reproduce the results in this paper.

\subsubsection*{Graphical display of simulated scenarios}

\end{document}